\documentclass[11pt]{article}

\usepackage[english]{babel}
\usepackage{sao1}
\usepackage{graphics}
\usepackage{epsfig}

\begin{document}

\title{$\tau$\,Sco: The discovery of the clones}
\author{V. Petit\inst{1} \and D.~L. Massa\inst{2} \and W.~L.~F. Marcolino\inst{3}
	\and G.~A. Wade\inst{4} \and R. Ignace\inst{5} \and the MiMeS Collaboration}
\institute{
Dept. of Geology \& Astronomy, West Chester University, West Chester, PA 19383, USA
\and
Space Telescope Science Institute, 3700 N. San Martin Drive, Baltimore, MD 21218, USA
\and
Observat\'orio Nacional-MCT, CEP 20921-400, S\~ao Crist\'ov\~ao, Rio de Janeiro, Brasil
\and
Dept. of Physics, Royal Military College of Canada, Kingston, Canada, K7K 4B4
\and
Dept. of Physics \& Astronomy, East Tennessee State University, Johnson City, TN 37614, USA
}

\maketitle 

\begin{abstract}

The B0.2 V magnetic star $\tau$\,Sco stands out from the larger population of massive magnetic OB stars due to its high X-ray activity and remarkable wind, apparently related to its peculiar magnetic field - a field which is far more complex than the mostly-dipolar fields usually observed in magnetic OB stars. $\tau$\,Sco is therefore a puzzling outlier in the larger picture of stellar magnetism - a star that still defies interpretation in terms of a physically coherent model.

Recently, two early B-type stars were discovered as $\tau$\,Sco analogues, identified by the striking similarity of their UV spectra to that of $\tau$\,Sco, which was - until now - unique among OB stars. We present the recent detection of their magnetic fields by the MiMeS collaboration, reinforcing the connection between the presence of a magnetic field and wind anomalies (Petit et al. 2010). We will also present ongoing observational efforts undertaken to establish the precise magnetic topology, in order to provide additional constrains for existing models attempting to reproduce the unique wind structure of $\tau$\,Sco-like stars. 

\end{abstract}

\section{The young magnetic B-type star $\tau$\,Sco}

Very little is known about the magnetic fields of hot, massive OB stars, due at least in part to the challenges of measurement. Even as a member of the elusive class of magnetic massive stars, the B0.2 V star $\tau$\,Sco is recognised to be a peculiar and outstanding object.

The magnetic field of $\tau$\,Sco is unique because it is structurally far more complex than the mostly-dipolar fields ($l=1$) usually observed in magnetic OB stars, with significant power in spherical-harmonic modes up to $l=5$ with a mean surface field strength of $\sim300$\,G (Donati et al. 2006)

$\tau$\,Sco also stands out from the crowd of early-B stars because of its stellar wind anomalies, as diagnosed through its odd UV spectrum (see Fig. 1).
The UV wind line morphology of normal, early B stars conforms to a 2-D spectral grid (Walborn et al. 1995).  Typically,  C\,\textsc{iv} strengthens with increasing temperature and luminosity. N\,\textsc{v} is at most a trace on the main sequence at spectral type B0\,V but strengthens with temperature and luminosity for more luminous stars.  
For stars with fixed C\,\textsc{iv} and N\,\textsc{v}, Si\,\textsc{iv} is strictly luminosity dependent and breaks the degeneracy. $\tau$\,Sco does not fit into this grid. It has strong N\,\textsc{v} indicating that it should be well above the main
sequence.  However, its C\,\textsc{iv} lines are only slightly stronger than typical and not distinctly wind-like, suggesting a near main sequence luminosity.  Finally, its Si\,\textsc{iv} profiles are unique, a bit stronger than typical class V stars, but unlike normal, early giants.  As a result, this stars lie outside of the normal classification grid, which suggests a more highly ionized outflow than typical. The hard X-ray emission of $\tau$\,Sco also suggests hot plasma, in excess of 10\,MK (Cassinelli et al. 1994).

\begin{figure}
\begin{center}
 \includegraphics[width=2.1in]{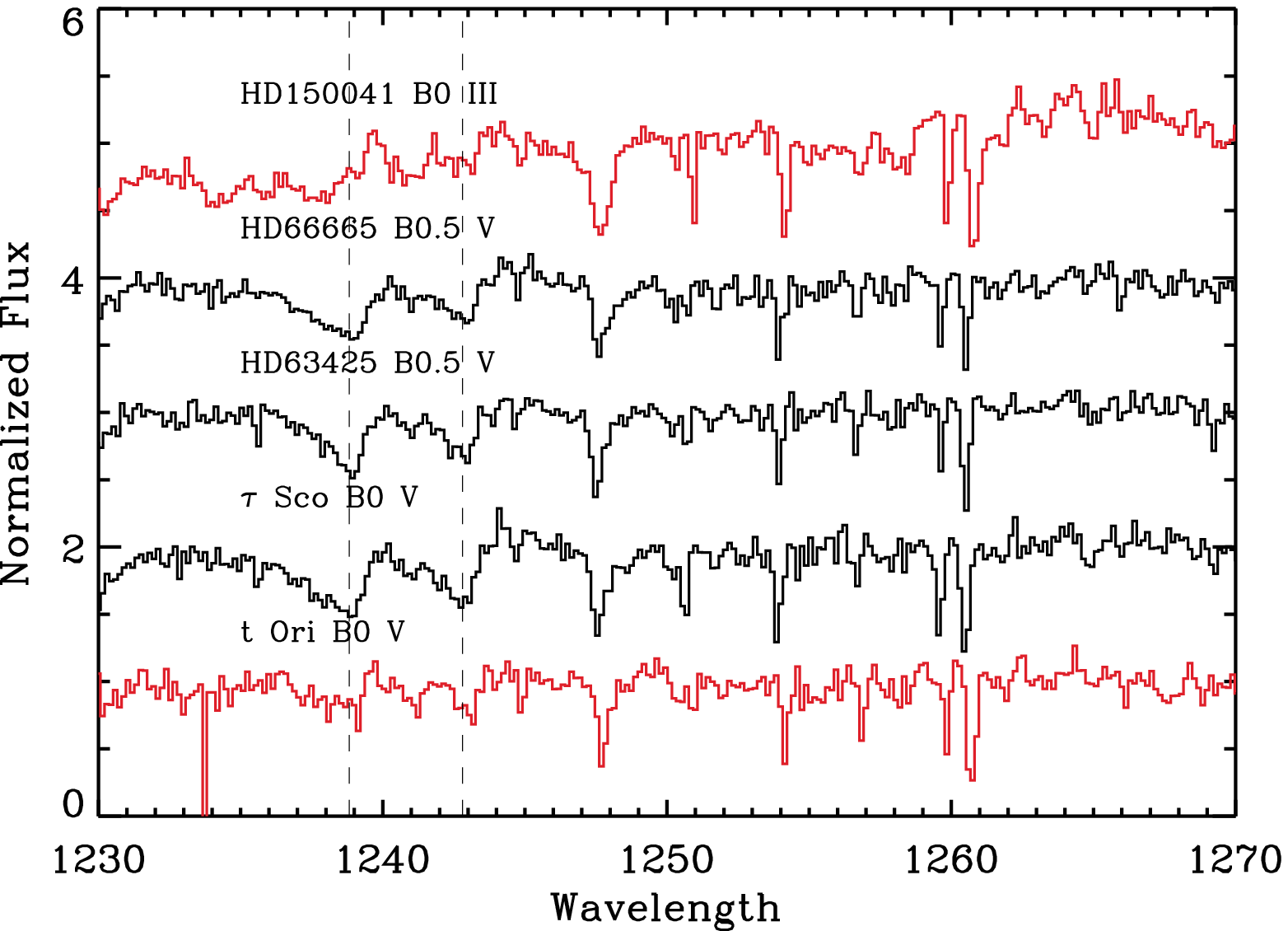} 
 \includegraphics[width=2.1in]{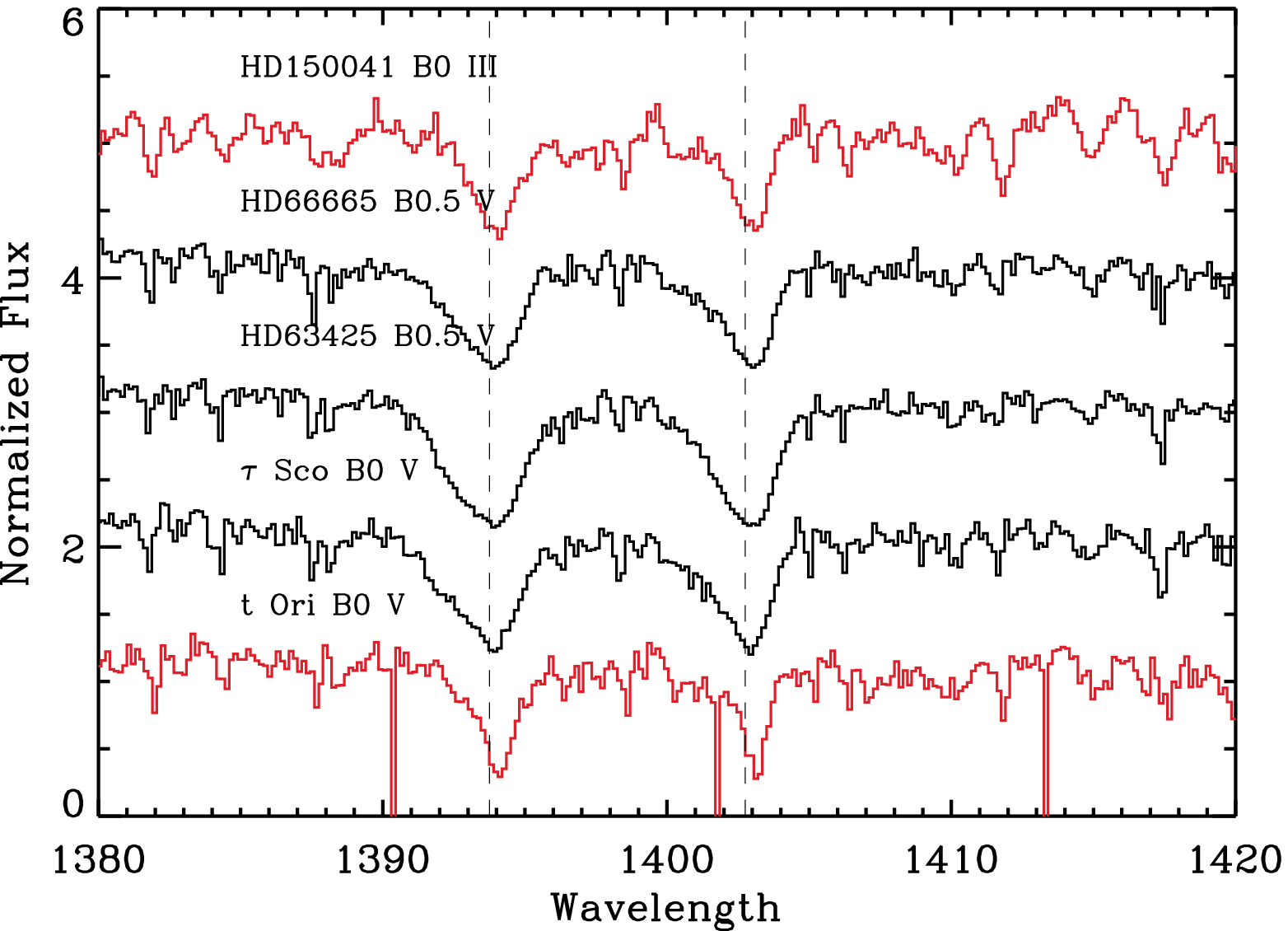} 
 \includegraphics[width=2.1in]{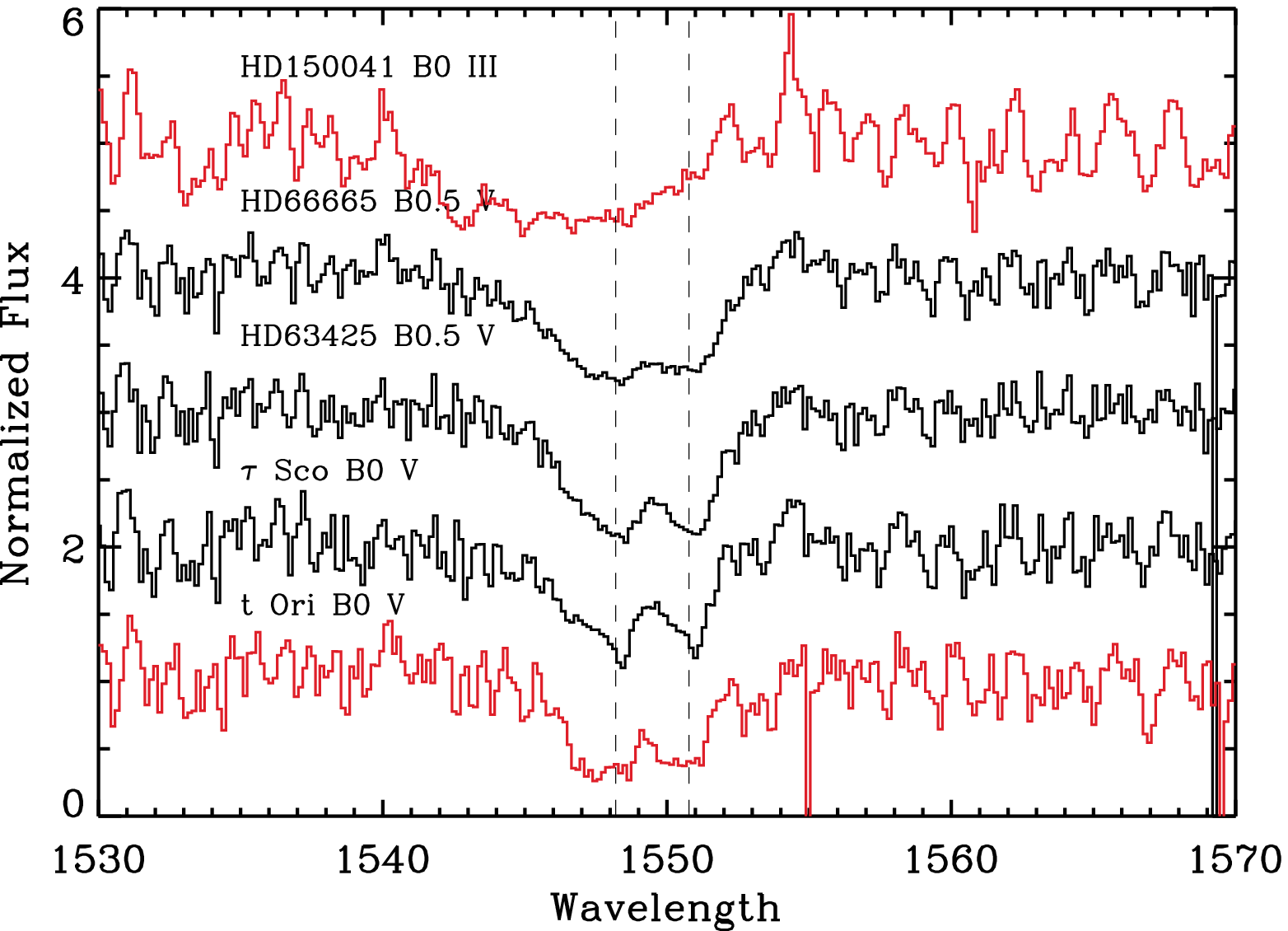}
 \caption{IUE spectra of $\tau$\,Sco and its analogues HD\,66665 and HD\,63425 (second to fifth spectra). For comparison, a typical spectrum for a B0 dwarf and a B0 giant is shown (bottom and top respectively). The dashed lines indicate the wind lines N\,\textsc{v}\,$\lambda \lambda 1239, 1243$, Si\,\textsc{iv}\,$\lambda\lambda 1393, 1403$ and C\,\textsc{iv}\,$\lambda \lambda 1548, 1550$. }
   \label{fig1}
\end{center}
\end{figure}

Interestingly, the wind lines of $\tau$\,Sco vary periodically with the star's 41\,d rotation period (Donati et al. 2006). Clearly the magnetic field exerts an important influence on the wind dynamics. What is not clear is whether the wind-line anomalies described above are a consequence of the unusual complexity of $\tau$\,Sco's magnetic field, a general consequence of wind confinement in this class of star, or perhaps even unrelated to the presence of a magnetic field.  

Because such wind anomalies have never been observed in any other star, magnetic or not, this issue has remained unresolved. 
The identification and analysis of additional stars with wind properties similar to $\tau$\,Sco would therefore represent an important step toward understanding the origin of these peculiarities. 

\section{The $\tau$\,Sco Clones}

We present two early B-type stars - HD\,66665 and HD\,63425 - that we identified to be the first $\tau$\,Sco analogues (Petit et al. 2010). These stars were first discovered by their UV spectra, which are strikingly similar to the UV spectrum of $\tau$\,Sco. 
The discovery of wind anomalies naturally led to an investigation of the magnetic properties of these stars by the Magnetism in Massive Stars (MiMeS) collaboration (Wade et al. 2010; this proceeding).

Spectropolarimetric observations of HD\,66665 and HD\,63425 were taken with ESPaSOnS at the Canada-France-Hawaii Telescope. We acquired 4 high-resolution, broad-band, intensity (Stokes I) and circular polarisation (Stokes V) for each star.

\section{Physical parameters}

In order to determine the stellar and wind parameters of HD\,66665 and HD\,63425 we used non-LTE model 
atmospheres from the \textsc{cmfgen} code (Hillier \& Miller 1998). The physical parameters, shown in Tab. 1, are similar to $\tau$\,Sco's. 
UV spectra from IUE were used to infer stellar wind properties. The absorption shown by C\,\textsc{iv} cannot be reproduced in detail by our models. On the other hand, we could achieve an acceptable fit for the Si\,\textsc{iv} and N\,\textsc{v} lines. The inclusion of X-rays in our \textsc{cmfgen} models was essential in producing the N\,\textsc{v} feature, since the wind structure is not sufficiently ionized without this component. 
We could derive an upper limit for the mass-loss rate from the emission part of the synthetic P-Cygni profile the C\,\textsc{iv} line and a rough estimate for the wind terminal velocity could be made. 

The mass-loss rate upper limits are much lower than the mass-loss rate assumed for $\tau$\,Sco ($6\times10^{-8}$\,M$_\odot$\,yr$^{-1}$; see Donati et al. 2006, and references therein). 
In fact, HD\,66665 and HD\,63425 should be considered as weak wind stars (Martins et al. 2005; Marcolino et al. 2009), since their expected mass-loss rate is about $10^{-8}$\,M$_\odot$\,yr$^{-1}$ (following the recipe provided by Vink et al. 2000). We therefore believe that the mass-loss rate found for $\tau$\,Sco in the literature has been over-estimated.

The apparent rotational velocity of these two stars is clearly low, but an accurate value is difficult to infer from the observed spectrum, due to turbulent broadening of comparable or larger magnitude.

\begin{table}
	\caption{Summary of stellar properties of $\tau$\,Sco, HD\,66665 and HD\,63425. }
	\begin{center}
	\begin{tabular}{l c c c}
	\hline
	 & $\tau$\,Sco$^1$ & HD\,66665 & HD\,63425 \\
	 \hline
	 Spec. type & B0.2\,V & B0.5\,V & B0.5\,V\\
	 $T_\mathrm{eff}$  (kK)& $31\pm1$ & $28.5\pm1.0$ & $29.5\pm1.0$ \\
	 $\log g$  (cgs)& $4.0\pm0.1$ & $3.9\pm0.1$ & $4.0\pm0.1$ \\
	 $R_\star$ (R$_\odot$)  & $5.6\pm0.8$ & $5.5\stackrel{+3.3}{_{-2.7}}$ & $6.8\stackrel{+4.8}{_{-2.0}}$ \\
 	 $\log(L_\star/\mathrm{L}_\odot)$ & $4.47\pm0.13$ & $4.25\stackrel{+0.40}{_{-0.60}}$ & $4.50\stackrel{+0.46}{_{-0.30}}$ \\
	 $M_\star$ (M$_\odot$) & $11\pm4$ &  $9\stackrel{+5}{_{-7}}$ & $17\stackrel{+34}{_{-9}}$ \\
	 $v\sin i$ (km\,s$^{-1}$) & $<13$ & $\la10$ & $\la15$ \\
	 $\dot{M}$ ($10^{-9}$\,M$_\odot$\,yr$^{-1}$)  & $61\stackrel{+10}{_{-2}}$ &  $< 0.45$  & $< 0.75$  \\
	$v_\infty$  (km\,s$^{-1}$)&    $\sim2000$        & $\sim1400$ & $\sim1700$ \\
	\hline
	\end{tabular}
	\end{center}
	$^1$ Parameters from Sim\'on-D\'iaz et al. (2006), and Mokiem et al. (2005) for $\dot{M}$ and $v_\infty$.
\end{table}

\section{Magnetic field}

\begin{figure}
\begin{center}
 \includegraphics[width=3in]{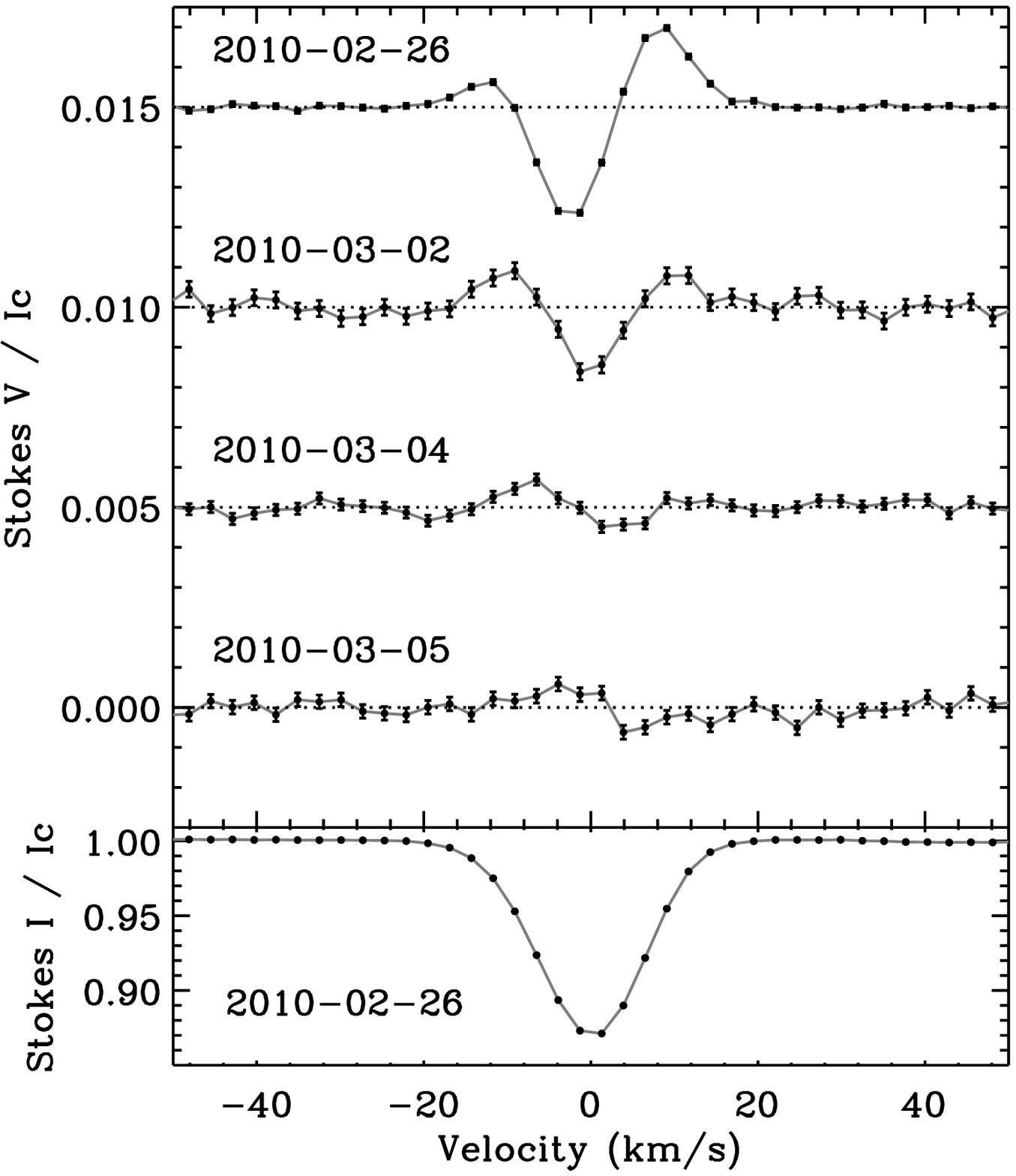} 
 \includegraphics[width=3in]{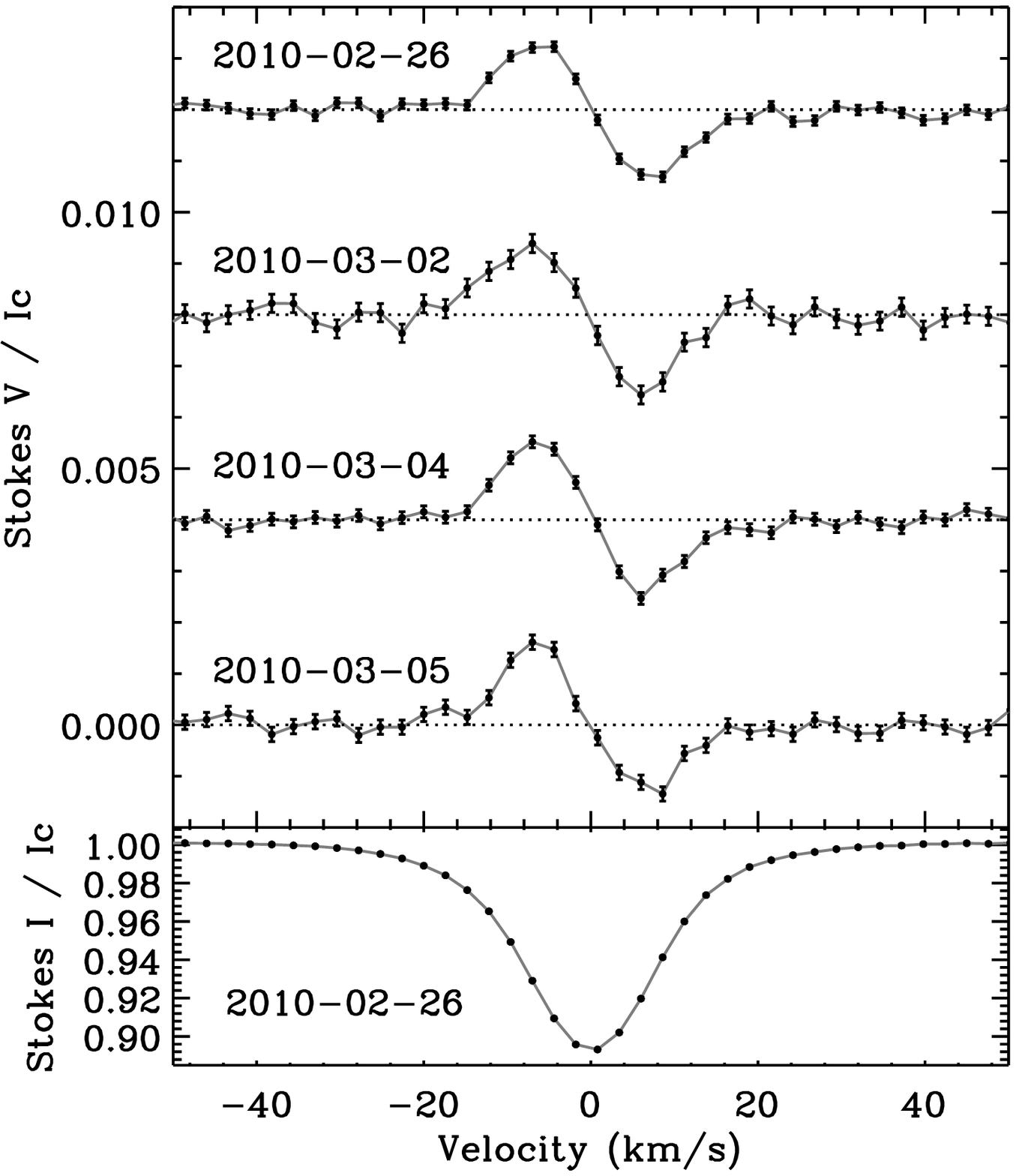} 
 \caption{Mean Stokes I absorption line profiles (bottom) and circular polarisation Stokes V profiles (top) of HD\,66665 (left) and HD\,63425 (right) obtained with ESPaDOnS. All LSD profiles have been scaled to correspond to a spectral line with weight corresponding to $d=0.2\,I_c$, $g=1.2$ and $\lambda_0=5000$\,\AA. }
\end{center}
\end{figure}

\begin{figure}
\begin{center}
 \includegraphics[width=3in]{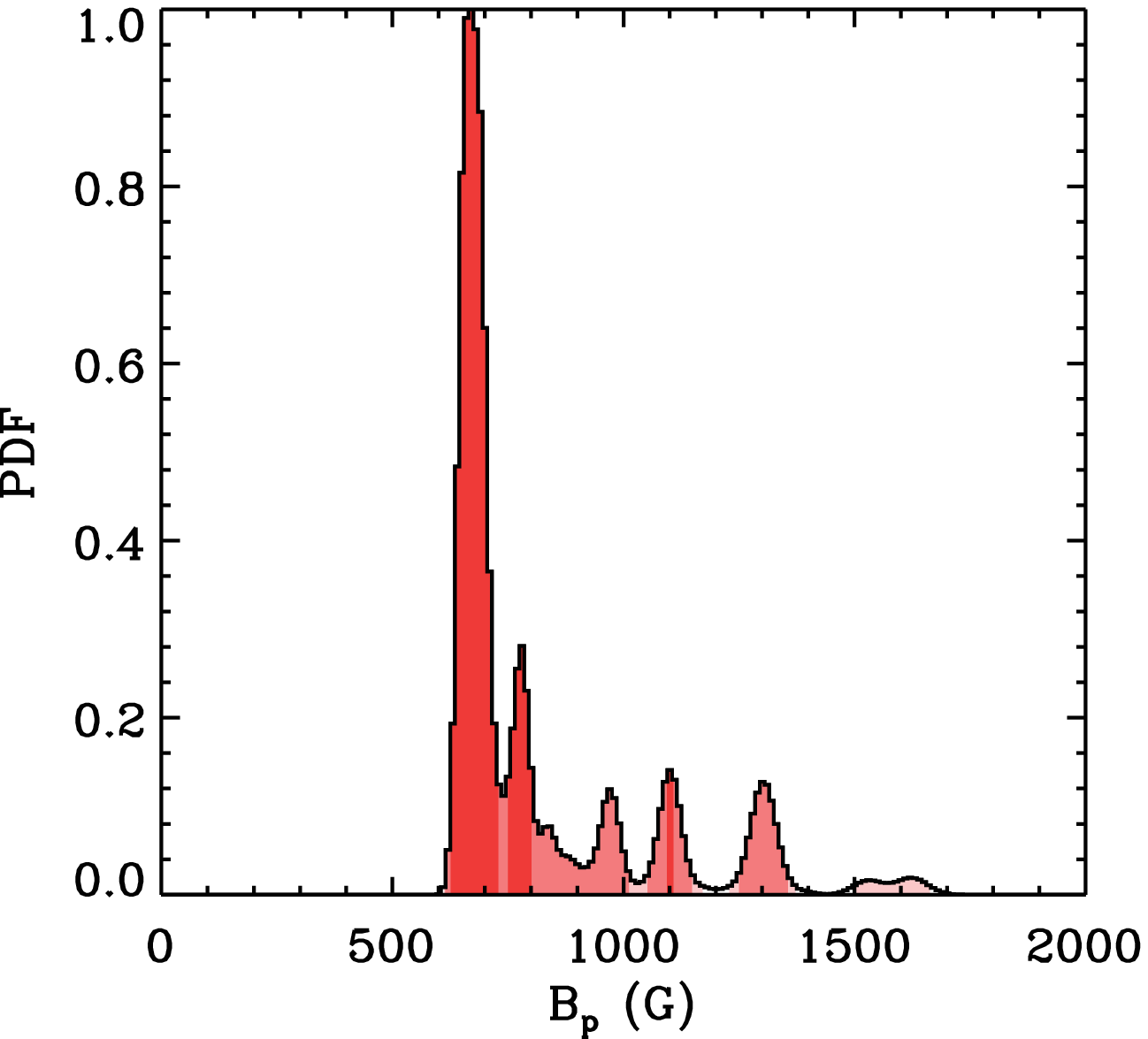} 
 \includegraphics[width=3in]{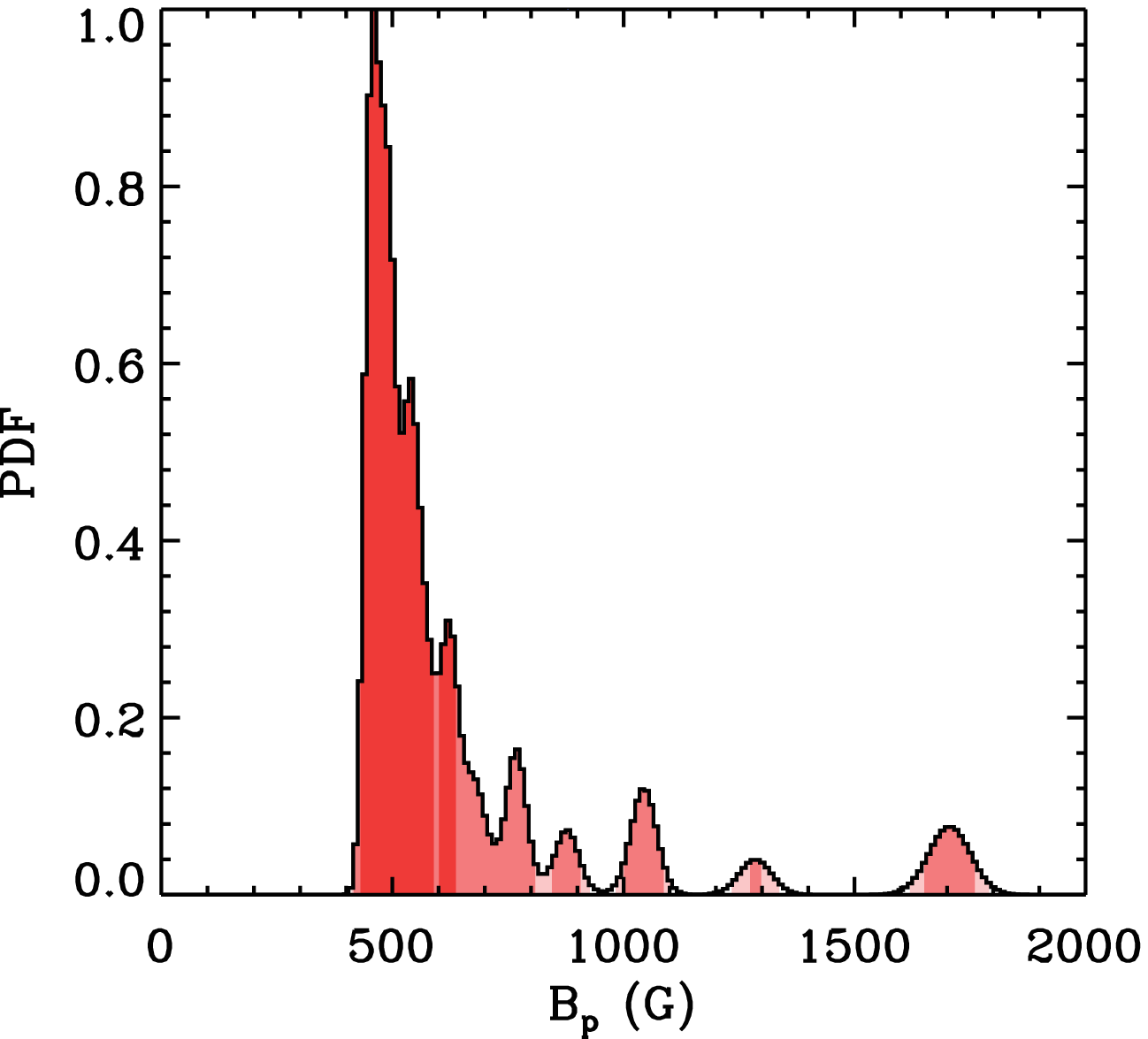} 
 \caption{Magnetic modelling of HD\,66665 (left) and HD\,63425 (right). For each star, we show the posterior probability density marginalized for the dipole strength $B_p$. The credible regions containing 68.3\%, 95.4\%, 99\% and 99.7\% of the probability are shaded, from dark to light colours respectively.}
\end{center}
\end{figure}

In order to increase the magnetic sensitivity of our data, we applied the Least-Squares Deconvolution (LSD) procedure, as described by Donati et al. (1997). This procedure enables the simultaneous use of many lines present in a spectrum to detect a magnetic field Stokes V signature.
Assumptions, limitations and numerical tests of this method are presented by Kochukhov et al. (2010).

All observations led to the detection of a magnetic signal (see Fig. 2). The same analysis was performed on the diagnostic null profiles, and no signal was detected.
As the exact rotation phases of our observations are not known, we used the method described by Petit et al. (2008; in prep), which compares the observed Stokes V profiles to a large grid of synthetic profiles, described by a dipole oblique rotator model. The model is parametrized by the dipole field strength $B_p$, the rotation axis inclination $i$ with respect to the line of sight, the positive magnetic axis obliquity $\beta$ and the rotational phase $\varphi$.  
Assuming that only $\varphi$ may change between different observations of a given star, the goodness-of-fit of a given rotation-independent ($B_p$, $i$, $\beta$) magnetic configuration can be computed to determine configurations that provides good posterior probabilities for all the observed Stokes V profiles, in a Bayesian statistic framework. 

The inclined dipole model can reproduce the current observations in a satisfactory fashion, although the $\chi^2_\mathrm{red}$ remains high for some observations (those with the highest s/n). Whether these deviations from the model come from artefacts of the LSD procedure, systematics introduced by our simple line profile modelling, or a more complex field, remains to be investigated once phase-resolved observations are obtained and a more detailed analysis can be performed. 

By treating any features that cannot be explained by the inclined dipole model as additional Gaussian noise, we can obtain a conservative estimate of the model parameters (see Fig. 3).
Our modelling of the LSD Stokes V profiles by an inclined dipole model results in field strengths\footnote{The surface-averaged modulus of a dipolar magnetic field is equal to 0.77 times the dipole polar field strength.} that are comparable with, or maybe slightly larger than, the mean surface field strength of $\tau$\,Sco. The exact value of the surface field strength is however highly dependent on the dipole geometry, and varies from 600\,G to 1.6kG for HD\,66665 and from 400\,G to 1.8\,kG for HD\,63425.

\section{Conclusion}

We have presented the characteristics of two stars - HD\,66665 and HD\,63425 - which we believe are analogues to the magnetic massive star $\tau$\,Sco. The UV spectra of these stars are similar to the once-unique spectrum of $\tau$\,Sco. 
We have shown that these three stars have similar fundamental properties, although the mass-loss rate values we estimate for HD\,66665 and HD\,63425 are lower than the value generally assumed for $\tau$\,Sco.
We have shown that these two stars host a magnetic field. Our modelling of the LSD Stokes V profiles by an inclined dipole model results in field strengths that are comparable with, or maybe slightly larger than, the mean surface field strength of $\tau$\,Sco. The current observations can be acceptably reproduced by the dipole model, although more phase-resolved observations are required in order to assess the potential complexity of their magnetic field, and verify if the wind anomalies are linked to the field complexity.

\end{document}